\documentstyle[epsf,amsfonts,floats,tighten,preprint,eqsecnum,prd,aps]{revtex}

\makeatletter

\def\invt{{{-1}^{\scriptstyle T}}}

\def\ddelta{\delta\hskip -.07em}

\newbox\sumbox
\setbox\sumbox=\hbox{$\sum$}
\def\notsum{\hbox to 0pt{\hskip .6\wd\sumbox minus 1 fil\mbox{\Large /}%
\hskip -.6\wd\sumbox minus 1 fil}\copy\sumbox}

\def\oldpaper{\cite{olum:paper1}}

\newenvironment{neqnarray}{\begingroup\arraycolsep 2pt\begin{eqnarray}}{\end{eqnarray}\endgroup}

\def\be{\@ifnextchar[{\be@label}{\begin{equation}}} 
\def\be@label[#1]{\begin{equation}\label{eqn:#1}}
\def\ee{\end{equation}}

\def\bea{\@ifnextchar[{\bea@label}{\begin{neqnarray}}} 
\def\bea@label[#1]{\begin{neqnarray}\label{eqn:#1}}
\def\eea{\end{neqnarray}}

\def\blea{\@ifnextchar[{\blea@label}{\begin{mathletters}\begin{neqnarray}}} 
\def\blea@label[#1]{\begin{mathletters}\label{eqn:#1}\begin{neqnarray}}
\def\elea{\end{neqnarray}\end{mathletters}}

\def\eqref#1{Eq.\ (\ref{eqn:#1})}
\def\eqsref#1{Eqs.\ (\ref{eqn:#1})} 
\def\Eqref#1{Equation (\ref{eqn:#1})}	
\def\Eqsref#1{Equations (\ref{eqn:#1})}
\def\figref#1{Fig.\ \ref{fig:#1}}

\def\secref#1{section \ref{sec:#1}}
\def\secsref#1{sections \ref{sec:#1}}
\def\oldsecref#1{section #1 of \oldpaper} 

\let\member=\in
\def\<{\langle }
\def\>{\rangle }
\def\Tr{\text{Tr}\,}

\def\freefree{{\text{free},\text{free}}}

\def\freeout{{\text{free},\text{out}}}
\def\freein{{\text{free},\text{in}}}
\def\gsfree{{\text{gs},\text{free}}}
\def\gsgs{{\text{gs},\text{gs}}}
\def\gsout{{\text{gs},\text{out}}}

\def\outfree{{\text{out},\text{free}}}
\def\outgs{{\text{out},\text{gs}}}
\def\outout{{\text{out},\text{out}}}
\def\outin{{\text{out},\text{in}}}
\def\infree{{\text{in},\text{free}}}

\def\inout{{\text{in},\text{out}}}
\def\inin{{\text{in},\text{in}}}
\def\midmid{{\text{mid},\text{mid}}}
\def\inmid{{\text{in},\text{mid}}}

\def\half{{1\over 2}}
\def\qtr{{1\over 4}}

\def\lowerleftnonzero#1{\arraycolsep 0pt
\renewcommand\arraystretch{0}
\left(\begin{array}{cc}
\noalign{\vspace{1ex}}
& \hspace{1em} 0 \hspace{1em} \\
\noalign{\vspace{1ex}}
\cline{1-1}
#1 \hspace{2pt} & \vline \hfill
\end{array}\right)}

\def\joinmatrix(#1,#2){
\left(\begin{array}{c|c}
   #1 & #2
   \end{array}\right)}

\def\sizedjoinmatrix(#1,#2)(#3,#4)(#5){
\left(\begin{array}{c|c}
	\noalign{\global\setbox\myarstrut=\box\@arstrutbox} 
	\matrixwidth{#3} & \matrixwidth{#4} \\ 
	 \copy\myarstrut #1 & #2 \\
	\end{array}\right)
\vrule width 0pt height 2\ht\myarstrut
\} \scriptstyle #5
}

\def\stackmatrix(#1,#2){
{\renewcommand\arraystretch{1.1}
\left(\begin{array}{c}
	#1 \\ \hline #2
	\end{array}\right)}}

\def\sizedstackmatrix(#1,#2)(#3)(#4,#5){
{\renewcommand\arraystretch{1.1}
\left(\begin{array}{c}
	\noalign{\global\setbox\myarstrut=\box\@arstrutbox} 
	\matrixwidth{#3} \\ 
	\copy\myarstrut #1 \\ \hline
	\copy\myarstrut #2
	\end{array}\right)
\vrule width 0pt height 3\ht\myarstrut
\hspace{-4pt}
\begin{array}{l}	
   \}\scriptstyle #3 \\
   \}\scriptstyle #4
\end{array}}}

\def\splitmatrix(#1,#2,#3,#4){
{\renewcommand\arraystretch{1.1} 
\left(\begin{array}{c|c}
	#1 & #2 \\ \hline
	#3 & #4
	\end{array}\right)
}}

\newbox\ulstrut
\savebox{\ulstrut}{\vrule width 0pt depth 10pt height 15pt}

\def\splitmatrixul(#1,#2,#3,#4){
{\renewcommand\arraystretch{1.1} 
\arraycolsep 2pt
\left(\begin{array}{c|@{\hspace{10pt}}c@{\hspace{10pt}}}
	#1 & #2 \\ \hline
	\copy\ulstrut #3 & #4
	\end{array}\right)
}}

\newbox\myarstrut
\newif\ifmatrixul		
\newlength{\myextrastrutlength}

\def\sizedsplitmatrix(#1,#2,#3,#4)(#5,#6)(#7,#8){
{\renewcommand\arraystretch{1.1} 
\ifmatrixul\arraycolsep 2pt\fi
\left(\begin{array}{c|\ifmatrixul @{\hspace{10pt}}c@{\hspace{10pt}}
		      \else c \fi}
	\noalign{\global\setbox\myarstrut=\box\@arstrutbox} 
	\matrixwidth{#5} & \matrixwidth{#6} \\ 
	 \copy\myarstrut #1 & #2 \\ \hline
	 \ifmatrixul \copy\ulstrut \else \copy\myarstrut \fi #3 & #4
	\end{array}\right)
}
\ifmatrixul
\setlength{\myextrastrutlength}{2\ht\myarstrut}
\addtolength{\myextrastrutlength}{\ht\ulstrut}
\vrule width 0pt height \myextrastrutlength
\else
\vrule width 0pt height 3\ht\myarstrut
\fi
\hspace{-4pt}
\begin{array}{l}	
   \}\scriptstyle #7 \\
   \ifmatrixul\copy\ulstrut\bigg\}\else\}\fi\scriptstyle #8
\end{array}
}

\def\matrixwidth#1{
  \vbox to 0pt{\vskip 0pt minus 1fil
    \hbox{$\scriptstyle #1$}\vskip 2pt}}

\def\sizedsplitmatrixul(#1,#2,#3,#4)(#5,#6)(#7,#8){
{\matrixultrue\sizedsplitmatrix(#1,#2,#3,#4)(#5,#6)(#7,#8)}}

\def\mypsfig#1#2{%
\begin{figure}
\begin{center}
\vspace{-1ex}
\leavevmode\epsfbox{#1.ps}
\vspace{-1.7ex}
\end{center}
\caption{#2}
\label{fig:#1}
\end{figure}}

\def\mypsfigoff#1#2{%
\begin{figure}
\begin{center}
\vspace{-1ex}
\leavevmode\hskip -15pt\epsfbox{#1.ps}
\vspace{-1.7ex}
\end{center}
\caption{#2}
\label{fig:#1}
\end{figure}}

\def\mypsfigr#1#2{\mypsfig{#1}{#2}\figref{#1}}
\def\mypsfigroff#1#2{\mypsfigoff{#1}{#2}\figref{#1}}

\newlength{\captionsize}
\long\def\@makecaption#1#2{
\vskip 10pt			
\setbox\@tempboxa\hbox{\sl #1: #2.}	
\hbox to\hsize{\hfil		
\ifdim \wd\@tempboxa >\captionsize	
\parbox{\captionsize}{\sl #1: #2.}	
\else\box\@tempboxa		
\fi
\hfil}}

\makeatother

\def\adag{a^\dagger}

\def\bdag{b^\dagger}

\def\calg{\cal G}

\def\calh{{\cal H}}

\def\Lin{L_\in}
\def\Linp{L_\in\hskip -1.7ex '\hskip 1.0ex}
\def\Linpsup#1{L_\in \hskip -1.7ex '\hskip .3ex^{(#1)}}
\def\Lout{L_\out}

\def\Nfree{N_\free}
\def\Ngs{N_\gs}
\def\Nin{N_\in}
\def\Ninn{{N_\in+1}}
\def\Nnorm{N_{\text{norm}}}
\def\Nout{N_\out}

\def\Px{{\bf P}_x}

\def\Pw{{\bf P}_w}
\def\bP{{\bf P}}
\def\Py{{\bf P}_y}
\def\Pz{{\bf P}_z}
\def\pop{{\Bbb P}}
\def\popsub{\pop}
\def\popex{\<\pop\>}
\def\popexsub{\popex}
\def\pwop{{\Bbb P}_w}
\def\pwopsub{(\pwop)}

\def\pzop{{\Bbb P}_z}

\def\popone{C}			
\def\pzopex{\<\pzop\>}
\def\pzopexsub{\pzopex}

\def\uvacinv{{U^\vac}^{\raise .5ex\hbox{$\scriptstyle -1$}}}

\def\bw{{\bf w}}

\def\wop{{\Bbb W}}
\def\wopsub{\wop}

\def\bx{{\bf x}}
\def\xop{{\Bbb X}}
\def\xopsub{\xop}
\def\xopex{\<\xop\>}
\def\xopexsub{\xopex}

\def\by{{\bf y}}

\def\bz{{\bf z}}
\def\zop{{\Bbb Z}}

\def\zopex{\<\zop\>}
\def\zopexsub{\zopex}

\def\rhofree{\tilde\rho}
\def\drhofree{\widetilde{\ddelta\rho}}

\def\free{{\protect\text{free}}}
\def\gs{{\text{gs}}}
\def\in{{\protect\text{in}}}
\def\mid{{\text{mid}}}
\def\out{{\text{out}}}
\def\rb{{\text{rb}}}
\def\vac{{\text{vac}}}

\begin{document}
\title{Entropy of very low energy localized states}

\author{Ken D.\ Olum\footnote{Email address: {\tt kdo@ctp.mit.edu}}
\footnote{New address: Institute of Cosmology, Department of Physics
and Astronomy, Tufts University, Medford, MA 02155}}

\address{Center for Theoretical Physics \\
Laboratory for Nuclear Science \\
and Department of Physics \\
Massachusetts Institute of Technology \\
Cambridge, Massachusetts 02139}

\date{hep-th/9709041, September 1997}

\maketitle

\begin{abstract}
We expand on previous work involving ``vacuum-bounded'' states, i.e.,
states such that every measurement performed outside a specified
interior region gives the same result as in the vacuum.  We improve
our previous techniques by removing the need for a finite outside
region in numerical calculations.  We apply these techniques to the
limit of very low energies and show that the entropy of a
vacuum-bounded state can be much higher than that of a rigid box state
with the same energy.  For a fixed $E$ we let $\Linp$ be the length of
a rigid box which gives the same entropy as a vacuum-bounded state of
length $\Lin$.  In the $E\rightarrow 0$ limit we conjecture that the
ratio $\Linp/\Lin$ grows without bound and support this conjecture
with numerical computations.
\end{abstract}


\section{Introduction}

In a previous paper \oldpaper\ we defined a ``vacuum-bounded state'' to be a
generalized state (i.e., density matrix) for which every operator
composed of fields at points outside a specified interior region has
the same expectation value as in the vacuum.  We used a vacuum-bounded
state as a model of the state resulting from the complete evaporation
of a black hole. We took the interior region to be a sphere of radius
$R$, the distance that information might have propagated after unknown
physics came into play in the evaporation process.

In the present paper we redo our analysis of maximum-entropy
vacuum-bounded states in a way that directly handles an infinite
outside region without the need to put the entire system in a box.
Using this technique we calculate the maximum entropies of
vacuum-bounded states with very low energies.  As in \oldpaper\
we do our calculations with a single scalar field in one dimension.
We argue that as the temperature of the system becomes very low, the
entropy is much larger than what one would have in a system with a
rigid boundary, and that this difference grows without bound as the
temperature goes to zero.

As in \oldpaper\ we approximate the continuum with a lattice of
harmonic oscillators.  Even with an infinite vacuum outside our
specified region, we can reduce the problem to a finite number of
degrees of freedom.  We then compute the solution numerically to
support, in this approximation, our claims of high entropy
at low energies.

We also use these techniques to recalculate the bounds we derived in
\oldpaper.  We obtain the same bound as in that paper, but now we work
directly with an infinite outside vacuum, which we could do only by
extrapolation in \oldpaper.

\section{Preliminaries}

We work with a one-dimensional scalar field with classical
Hamiltonian
\be
H = \half \int_0^L \left[\pi(x)^2 
 + \left({d\phi\over dx}\right)^2\right] dx\,.
\ee
We let $\Lin$ be the size of the inside region.  We will initially put
the system in a box of length $L$, so the outside region stretches
from $x=\Lin$ to $x=L$, but we will shortly take $L$ to $\infty$.

Our goal is to find a density matrix $\rho$ such that if $O_\out$ is
any operator composed of fields at points $x_i > \Lin$, then
\be[const1]
\Tr\rho O_\out = \<0|O_\out|0\>\,.
\ee
From \oldpaper\ we expect $\rho$ to be a Gaussian of the form
\be
\rho \propto e^{-\beta H+\sum f_\alpha O_2^\alpha}
\ee
where $O_2^\alpha$ ranges over the quadratic operators
$\phi(x)\phi(y)$ and $\pi(x)\pi(y)$ with $x,y > \Lin$.  The
coefficients $f_\alpha$ must then be chosen so that $\rho$ satisfies
\eqref{const1}.

In discrete form, we have \oldpaper
\be[hdisc]
H = \half\left(\Px \cdot \Px + \bx \cdot K \bx\right)\,.
\ee
with
\be
K = \left(\matrix{2g & -g & 0 & 0 & \hspace{5pt}\cdot\hspace{5pt} \cr
                    -g & 2g & -g & 0 & \cdot \cr
                    0 & -g & \cdot & \cdot & \cdot \cr
                    0 & 0 & \cdot & \cdot & \cdot \cr
                    \cdot & \cdot & \cdot & \cdot & \cdot \cr
                    }\right)
\ee
where $g = 1/L_1^2$ and $L_1 = L/(N+1)$ is the lattice spacing.

We will take $\Nin \approx \Lin/L_1$ of the oscillators to represent
the inside region, and $\Nout = N - \Nin$ to represent the outside
region.  \Eqref{const1} then becomes
\blea[rhoconstraints]
\Tr\rho x_i x_j & = & \<0|x_i x_j|0\>\label{eqn:xrhoconstraint}\\ 
\Tr\rho P_i P_j & = & \<0|P_i P_j|0\>\label{eqn:Prhoconstraint}\,,
\elea
where $i$ and $j$ run over the oscillators which represent the outside
region.  We require also that $\rho$ satisfy an energy constraint,
\be
\Tr\rho H = E_0\label{eqn:Hrhoconstraint}\,.
\ee

We will define matrices $\xop$ and $\pop$ whose elements are
the quadratic operators via
\blea
\xopsub_{\mu\nu} &=& x_\mu x_\nu\\
\popsub_{\mu\nu} &=& P_\mu P_\nu
\elea
so that \eqsref{xrhoconstraint} and (\ref{eqn:Prhoconstraint})
become
\blea[opconstraints]
\Tr\rho\xop_\outout &=& \<0|\xop_\outout|0\>\\
\Tr\rho\pop_\outout &=& \<0|\pop_\outout|0\>\,.
\elea
We expect $\rho$ to have the form
\be[rhoform]
\rho \propto e^{-\beta\left(H + f_{ij}x_i x_j +g_{ij}P_i
P_j\right)}\,.
\ee

\section{Reduction}
\label{sec:reduction}

\Eqref{rhoform} has $\Nout(\Nout+1)$ degrees of freedom in the $f$ and $g$
parameters.  We would like to reduce these degrees of freedom to a
number which does not depend on $\Nout$, but only on $\Nin$, so that
we can take $\Nout$ to infinity and still have a well-defined problem.
To make this reduction we will describe our system by a new set of
modes, such that the number of modes which can be excited in any
$\rho$ which satisfies \eqref{opconstraints} depends only on $\Nin$.

\subsection{A different description of the vacuum}

\label{sec:oldgroundstate}

We will start by looking at the vacuum state.  We can make a change of
coordinate to put the Hamiltonian of \eqref{hdisc} in diagonal form.
Let $Z$ be a matrix whose columns are the eigenvectors of $K$, $K = Z
\Omega_0^2 Z^{-1}$, with the normalization
\be[Znorm]
Z\Omega_0 Z^T = I \qquad\hbox{and}\qquad Z^\invt \Omega_0 Z^{-1} = K\,.
\ee
Define new coordinates $\bz$ via
$\bx = Z \bz$ and $\Px = Z^\invt \Pz$.  In these coordinates,
\be
H = \half \sum_\alpha \omega^{(0)}_\alpha 
\left(P_{z_\alpha}^2 + z_\alpha^2\right)\,.
\ee
The vacuum is the ground state of this Hamiltonian.  We can define
raising and lowering operators
\blea
a_\alpha & = & {1\over\sqrt{2}}\left(z_\alpha+iP_{z_\alpha}\right)\\
\adag_\alpha & = & {1\over\sqrt{2}}\left(z_\alpha-iP_{z_\alpha}\right)\\
H & = & \sum_\alpha \omega^{(0)}_\alpha \left(\adag a + \half\right)\,.
\elea
The vacuum is the state $|0\>$ annihilated by all the $a_\alpha$.
It is straightforward to write the expectation values in the vacuum
state,
\blea
\zopexsub_{\alpha\beta} \equiv \<0|z_\alpha z_\beta|0\>= \half\delta_{\alpha\beta}\\
\pzopexsub_{\alpha\beta} \equiv \<0|P_{z_\alpha} P_{z_\beta}|0\>
= \half\delta_{\alpha\beta}
\elea
so
\blea[xxppdef]
\<0|\xop|0\> & = & Z \<0|\zop|0\> Z^T = \half Z Z^T\\
\<0|\pop|0\> & = & Z^\invt \<0|\pzop|0\> Z^{-1}= \half Z^\invt Z^{-1}
= {1\over 4}(\<0|\xop|0\>)^{-1}\label{eqn:xoppinv}\,.
\elea

There can be many different
Hamiltonians that have the same ground state.  If we consider
\be
H' = \half\left(\Px \cdot T' \Px + \bx \cdot K' \bx\right)
\ee
with $T'$ and $K'$ some coupling
matrices, we can follow the
above derivation to get a normal mode matrix $Y$ and some
frequencies $\Omega$ with
\blea
Y \Omega Y^T & = & T'\\
Y^\invt \Omega Y^{-1} & = & K'\\
\bx & = & Y \by\\
\Px & = & Y^\invt \Py\\
H' & = & \half \sum_\beta \omega_\beta 
\left(P_{y_\beta}^2 + y_\beta^2\right)\,.
\elea
We can define raising and lowering operators for these modes,
\blea
b_\beta & = & {1\over\sqrt{2}}\left(y_\beta+iP_{y_\beta}\right)\\
\bdag_\beta & = & {1\over\sqrt{2}}\left(y_\beta-iP_{y_\beta}\right)\\
H' & = & \sum_\beta \omega_\beta \left(\bdag b + \half\right)\,.
\elea

Now $\by = Y^{-1} \bx = Y^{-1} Z \bz = W^{-1} \bz$ where
\be
W \equiv Z^{-1} Y\,.
\ee
Similarly, $\Py = Y^T\Px = Y^T Z^\invt \Pz = W^T \Pz$.
Consequently we can define a Bogoliubov transformation,
\bea
b_\beta & = & {1\over\sqrt{2}}\left(W^{-1}_{\beta\alpha}z_\alpha
+ iW^T_{\beta\alpha}P_{z_\alpha}\right)\nonumber\\
& = & \half\left(W^{-1}_{\beta\alpha}(a_\alpha + \adag_\alpha)
+ W^T_{\beta\alpha}(a_\alpha - \adag_\alpha)\right)\nonumber\\
& = & \half\left((W^{-1}+W^T)_{\beta\alpha}a_\alpha
+ (W^{-1} - W^T)_{\beta\alpha}\adag_\alpha\right) \,.
\eea
For $H$ and $H'$ to have the same vacuum we require that the $b_\beta$
depend only on the $a_\alpha$ and not on the $\adag_\alpha$, which is
to say that $W^{-1} = W^T$, i.e.\ that $W$ is a unitary matrix. With
$W$ unitary, $Y Y^T = Z Z^T$ so the vacuum expectation values of
\eqref{xxppdef} have the same values expressed in terms of $Y$ as they
had in terms of $Z$.

\subsection{Modes that remain in the ground state}

Now consider a unitary matrix $W$ and let $Y = Z W$ as in the last
section.  Suppose we can find $W$ such that $Y$ has the following
property:
\begin{quote}
The $N$ modes can be divided into $\Ngs > 0$ ``ground state'' modes
and $\Nfree \equiv N - \Ngs$ ``free'' modes such that for all $a \le
\Nin$ and for all $\beta > \Nfree$, $Y_{a\beta} = 0$ and
$Y^{-1}_{\beta a} = 0$.
\end{quote}
That is to say $Y$ and $Y^{-1}$ will have the form
\be[Yform]
Y = \sizedsplitmatrix(Y_\infree, 0, Y_\outfree, Y_\outgs)(\Nfree, \Ngs)(\Nin, \Nout)
\ee
and
\be[Yinvform]
Y^{-1} = \sizedsplitmatrix(Y^{-1}_\freein, Y^{-1}_\freeout, 0,
			Y^{-1}_\gsout)(\Nin, \Nout)(\Nfree,\Ngs)
\,.
\ee

If this is the case, then for $\beta > \Nfree$,
$b_\beta = \left(Y^{-1}_{\beta\gamma}x_\gamma
+ i Y_{\gamma\beta} P_{x_\gamma}\right)/\sqrt{2}$ depends only on outside
operators $x_i$ and $P_{x_i}$.  In any vacuum-bounded state,
regardless of entropy considerations, $\<x_i x_j\>$ and $\<P_iP_j\>$
have their vacuum values.  Consequently, for all $\beta > \Nfree$, if
$\rho$ describes a vacuum-bounded state then $\Tr\rho\bdag_\beta
b_\beta = \<0|\bdag_\beta b_\beta|0\> = 0$.  The vacuum-bounded
constraint forces modes $\beta > \Nfree$ to be in their ground states.
These modes will not contribute to the calculation of the
maximum-entropy vacuum-bounded state.

How many such modes can exist?  Let $W_\beta$ denote a column of $W$
with $\beta > \Nfree$ and let $Z^a$ denote a row of $Z$ with $a
\le \Nin$.  Similarly let $(Z^{-1})_a$ denote a column of
$Z^{-1}$.  Since $Y = Z W$, $Y_{a \beta} = 0$ whenever $W_\beta$
is orthogonal to $Z^a$.  Similarly $Y^{-1} = W^T Z^{-1}$ so
$Y^{-1}_{\beta a} = 0$ whenever $W_\beta$ is orthogonal to
$(Z^{-1})_a$.  Since $W$ is unitary, the $W_\beta$ must also be
orthogonal to each other.  Thus there are $\Ngs$ columns of $W$ which have
to be orthogonal to $\Nin$ rows of
$Z$, to $\Nin$ columns of $Z^{-1}$, and to each other.  Since
there are $N$ components in a column of $W$ it can be orthogonal in
general to at most $N-1$ other vectors.  Thus $\Ngs$ is limited by
$N-1 = 2\Nin + \Ngs -1$, or $\Ngs = N - 2\Nin = \Nout - \Nin$.  Thus
whenever $\Nout > \Nin$ there will be $\Ngs = \Nout - \Nin$ modes that are
forced to remain in the ground state.

These conditions determine the columns $W_\beta$ with $\beta >
2\Nin$ up to a unitary transformation on these columns alone, and
likewise the remaining columns are determined up to a unitary matrix
which combines them.

\subsection{Density matrix and entropy}

Let $\calh$ be the Hilbert space of states of our system.  We can
write $\calh = \tilde\calh \otimes \calg$ where $\tilde
\calh$ is the Hilbert space of states of the
``free'' modes and $\calg$ is the Hilbert space of states of the
``ground state'' modes.  Let $|0\>_\gs\member\calg$ denote the ground
state of system $\calg$.  For any operator $A$ we can define an operator
$\tilde A$ that acts on $\tilde\calh$ via
$\<\tilde\alpha|\tilde A|\tilde\beta\> =
\<\tilde\alpha\otimes 0_\gs| A | \tilde\beta\otimes 0_\gs\>$ for all
$\tilde\alpha,\tilde\beta\member\tilde\calh$.

Let $\rho$ describe a vacuum-bounded state and let $\rho_\gs =
\Tr_\free\rho$.  Then $\Tr\rho_\gs\bdag_\beta b_\beta = 0$.  This
defines the vacuum state, so $\rho_\gs = |0\>_\gs\<0|_\gs$ and thus
 \be
\rho = \rhofree \otimes\left(|0\>_\gs\<0|_\gs\right)\,.
\ee
If we write $\rhofree$ in diagonal form,
\be
\rhofree = \sum_\alpha P_\alpha |\tilde\alpha\>\<\tilde\alpha|\,,
\ee
then
\be
\rho = \sum_\alpha P_\alpha |\tilde\alpha\otimes 0_\gs\>
\<\tilde\alpha \otimes 0_\gs|\,.
\ee
The entropy is
\be
S = -\Tr\rho\ln\rho = - \sum_\alpha s(P_\alpha) = - \Tr\rhofree\ln\rhofree\,.
\ee
Since any vacuum-bounded $\rho$ has this form, a variation of $\rho$
that preserves the constraints must also have this form,
\be
\ddelta\rho = \drhofree \otimes \left(|0\>_\gs\<0|_\gs\right)\,.
\ee
If $S = -\Tr(\rho + t\ddelta\rho) \ln(\rho+t\ddelta\rho)
= -\Tr(\rhofree + t\drhofree) \ln(\rhofree+t\drhofree)$, then
\be
{dS\over dt} = - \Tr\drhofree\ln\rhofree\,.
\ee
If $\rho$ maximizes $S$ subject to the constraints, then we must have
$\Tr\drhofree\ln\rhofree$ for any variation $\drhofree$ that
preserves the constraints, i.e.\ for which $\Tr\drhofree = 0$ and
\be
\Tr \ddelta\rho H = \Tr \ddelta\rho x_i x_j = \Tr  \ddelta\rho P_i P_j = 0\,.
\ee
Now for any $A$, $\Tr\ddelta\rho A = \Tr (\drhofree\otimes(|0_\gs\>\<0_\gs|))A$.
When we take the trace we only need to sum over states of the form
$|\tilde\alpha\otimes 0_\gs\>$.  Thus
\be[opistilde]
\<A\> = \Tr\ddelta\rho A = \sum_{\tilde\alpha\tilde\beta}
\<\tilde\alpha|\drhofree|\tilde\beta\>
\<\tilde\beta\otimes 0_\gs|A|\tilde\alpha\otimes 0_\gs\>
= \Tr\drhofree\tilde A = \<\tilde A\>\,.
\ee

Thus we are looking for $\rhofree$ that maximizes
$S=-\Tr\rhofree\ln\rhofree$ subject to the constraints
$\Tr\drhofree\tilde H = \Tr \drhofree \widetilde{x_i x_j} = \Tr
\drhofree\widetilde{P_i P_j} = 0$, where $i$ and $j$ range over the
outside oscillators.

From \eqref{rhoform} we expect $\rhofree$ to have the form
\be[ourrhoform]
\rhofree \propto e^{-\beta\left(\tilde H + f_{ij}\widetilde{x_i x_j}
+g_{ij}\widetilde{P_i P_j}\right)}\,.
\ee
As in \oldpaper\, we can write this
\be
\rho \propto e^{-\beta H'}\,,
\ee
where $H'$ is a fictitious Hamiltonian for these oscillators,
\be[Hprimeform]
H' = \tilde H + f_{ij}\widetilde{x_i x_j} +g_{ij}\widetilde{P_i P_j}\,.
\ee

\subsection{New coordinates}
\label{sec:startreduction}

We would now like to introduce new coordinates $\bw$ as follows: The
first $\Nin$ $\bw$ coordinates will be the inside oscillator
coordinates, $\bw_\in = \bx_\in$.  The last $N_\gs$ coordinates are
the ground state normal modes, $\bw_\gs = \by_\gs$.  The remaining
$\Nin$ coordinates can be any coordinates that are independent of
those specified so far; we will make a particular choice later.

To do this, we proceed as follows:  Using \eqsref{Yform} and
(\ref{eqn:Yinvform}) we can write out
\be
I = Y^{-1} Y =
\splitmatrix(Y^{-1}_\freein Y_\infree+Y^{-1}_\freeout Y_\outfree,
	Y^{-1}_\freeout Y_\outgs,Y^{-1}_\gsout Y_\outfree,
	Y^{-1}_\gsout Y_\outgs)\,.
\ee
In particular, $Y^{-1}_\gsout Y_\outgs = I$.  We would like to extend
$Y^{-1}_\gsout$ and $Y_\outgs$ into square matrices $R$ and $R^{-1}$ with
\be
R = \sizedjoinmatrix(D, Y_\outgs)(\Nin, \Ngs)(\Nout)
\ee
and
\be
R^{-1} = \sizedstackmatrix(D', Y^{-1}_\gsout)(\Nout)(\Nin,\Ngs)\,.
\ee
This means that we must find $D$ and $D'$ such that
\blea[dconstraints]
Y^{-1}_\gsout D & = & 0\\
D' Y_\outgs & = & 0\\
D' D & = & I\,.
\elea
There are many possible choices of $D$ and $D'$ that satisfy
\eqsref{dconstraints}.  Here we proceed as follows:  Let
\blea
Z & = & \stackmatrix(Z_\in, Z_\out)\\
Z^{-1} & = & \joinmatrix(Z^{-1}_\in, Z^{-1}_\out)
\elea
so that $Z_\in Z^{-1}_\in = I$, $Z_\out Z^{-1}_\out = I$ and
$Z^{-1}_\in Z_\in + Z^{-1}_\out Z_\out = I$.
Now let
\blea[ddbardef]
\bar D & = & \half Z_\out Z_\in^T = \<0|\xop|0\>_\outin\\
\bar D' & = & \half Z_\in^\invt Z^{-1}_\out = \<0|\pop|0\>_\inout\,,
\elea
let $A$ and $B$
be $\Nin \times \Nin$ matrices with 
\be[abinv]
A B = (\bar D' \bar D)^{-1}
\ee
and let
\blea[ddprime]
D & = & \bar D A\\
D' & = & B \bar D'\,.
\elea
We also divide
\be
W = \joinmatrix(W_\free, W_\gs)\,.
\ee
Since $Y=ZW$ and $Y^{-1} = W^T Z^{-1}$, we have
\blea[zw-zero]
Z_\in W_\gs &= & 0\\
W_\gs^T Z^{-1}_\in & = & 0\,.
\elea
Thus
\nopagebreak
\be
D'Y_\outgs \propto B Z_\in^\invt Z^{-1}_\out Z_\out W_\gs
= B Z_\in^\invt W_\gs - B Z_\in^\invt Z^{-1}_\in Z_\in W_\gs
= 0
\ee
by \eqsref{zw-zero} and their transposes.  Similarly
\be
Y^{-1}_\gsout D \propto W_\gs^T Z^{-1}_\out Z_\out Z_\in^T A
= W_\gs^T Z_\in^T A - W_\gs^T Z^{-1}_\in Z_\in Z_\in^T A = 0\,.
\ee
From \eqref{abinv} we find $D'D = I$.  Thus the matrices $D$ and $D'$
satisfy \eqsref{dconstraints}.  We still have the freedom of
choosing the matrix $A$ arbitrarily.

Now let
\be
Q = \sizedsplitmatrix(I,0,0,R)(\Nin,\Nout)(\Nin,\Nout)
\ee
and define $\bw$ by $\bx = Q\bw$ so $\Px = Q^\invt \Pw$.  Then
$\bw_\in = \bx_\in$ and $\bw_\gs = \by_\gs$ as desired.

\label{sec:xfromw}

\subsection{Reduced operators}

We would like to recast our problem in terms of $\bw_\free$, the the
first $\Nfree$ $\bw$ coordinates.  First we look at the operators $x_\mu
x_\nu = \xopsub_{\mu\nu}$ and $P_\mu P_\nu = \popsub_{\mu\nu}$.  If we write
\blea
Q &=& \joinmatrix(Q_\free, Q_\gs)\\
Q^{-1} &=&  \stackmatrix(Q^{-1}_\free, Q^{-1}_\gs)\,,
\elea
we have $\bx = Q \bw
= Q_\free \bw_\free + Q_\gs \bw_\gs$ and
$\bP = Q^\invt \Pw = Q^\invt_\free (\Pw)_\free + Q^{-1}_\gs
(\Pw)_\gs$, so
\blea
\xop &=& Q_\free \wopsub_\freefree Q_\free^T + Q_\gs \wopsub_\gsgs
Q_\gs^T\\
\pop &=& Q^\invt_\free \pwopsub_\freefree Q^{-1}_\free
+ Q^\invt_\gs  \pwopsub_\gsgs Q^{-1}_\gs\,.
\elea
Now we form the reduced operators $\widetilde{x_\mu x_\nu}$ and
$\widetilde{P_\mu P_\nu}$.  Since $\bw_\gs = \by_\gs$ and the $\by_\gs$
modes are in the ground state by definition, $\widetilde\wopsub_\gsgs =
\widetilde\pwopsub_\gsgs = (1/2) I$, and so
\blea[expandoptilde]
\tilde\xop &=& Q_\free \wopsub_\freefree Q_\free^T + \half Q_\gs Q_\gs^T\\
\tilde\pop &=& Q^\invt_\free \pwopsub_\freefree Q^{-1}_\free
+ \half Q^\invt_\gs Q^{-1}_\gs\,.
\elea
In each case the second term is just a constant.
Now
\be
H = \half\left(P_\mu P_\mu + K_{\mu\nu} x_\mu x_\nu\right)
= \half\Tr\left(\pop + K\xop\right)
\ee
where the trace is over the oscillator indices. Thus
\bea[tildeHdef]
\tilde H & = & \half\Tr\left(Q^\invt_\free \pwopsub_\freefree Q^{-1}_\free
+ K Q_\free \wopsub_\freefree Q_\free^T\right)+\text{const}\nonumber\\
& = & \half\Tr\left(\tilde T \pwopsub_\freefree + \tilde K
\wopsub_\freefree \right) + \text{const}
\eea
where
\blea
\tilde T &=& Q^{-1}_\free Q^\invt_\free = 
\sizedsplitmatrix(I, 0, 0, D' D'^T)(\Nin,\Nin)(\Nin,\Nin)\\
\tilde K &=& Q_\free^T K Q_\free =
   \sizedsplitmatrix(K_\inin, K_\inout D, D^T K_\outin, D^T K_\outout D)(\Nin,\Nin)(\Nin,\Nin)
\label{eqn:ktildedef}\,.
\elea

The constant term in \eqref{tildeHdef} is
\be
\half\Tr\left(K Q_\gs Q_\gs^T+Q^\invt_\gs Q^{-1}_\gs\right)
= \half\Tr\left(Y_\outgs^T K_\outout Y_\outgs + Y^\invt_\gsout
Y^{-1}_\gsout\right)\,.
\ee
It depends on the ground state modes only and is just part of
the zero-point energy.  It will be the same in the vacuum and in a
vacuum-bounded state.  Thus if instead of \eqref{tildeHdef} we use
\be[tildeHdef-noconst]
\tilde H = \half\Tr\left(\tilde T \pwopsub_\freefree + \tilde K
\wopsub_\freefree \right)
\ee
we are just shifting $\tilde H$ by a constant term and thus changing
the zero-point energy.

Now we would like to make this reduced system look as much as possible
like the system we started with.  Let $B_1$ be some matrix such that
$B_1^T B_1 = (\bar D' \bar D'^T)^{-1}$, let $B_2$ be a unitary matrix
to be determined, and let $B = B_2 B_1$.  Then $D' D'^T = I$, so
$\tilde T = I$.  This gives $A = A_1 A_2$ where $A_1 = (\bar D' \bar
D)^{-1} B_1^{-1}$ and $A_2 = B_2^T$.  Let $D_1 = \bar D A_1$ and let
\be
\tilde K_1= \splitmatrix(K_\inin, K_\inout D_1, D_1^T K_\outin, D_1^T K_\outout D_1)\,.
\ee

Now
\be
K_\inout = \lowerleftnonzero{-g}\,,
\ee
so $K_\inout D_1$ is nonzero only in the last row.  Thus the last
$\Nin+1$ rows and columns of $\tilde K_1$ look like
\be
\sizedsplitmatrixul(2g, ?, ?, ?)(1,\Nin)(1,\Nin)\,.
\ee
A matrix of size $(\Ninn)\times(\Ninn)$ can be put in tridiagonal form by
a unitary transform of the form
\be
\sizedsplitmatrixul(1, 0, 0, U)(1,\Nin)(1,\Nin)
\ee
which can be constructed, for example, using the Householder process.
We will use this to choose $A_1 = U$, so that $\tilde K$ will be
tridiagonal.  For each off-diagonal element in the resulting
tridiagonal matrix there is a choice of sign, and we will choose them
all to be negative.  Thus in the $\bw$ coordinates, each oscillator
has unit mass and is coupled only to its neighbors.  This completely
specifies the matrices $A$ and $B$ and thus $D$ and $D'$.  Note that
$D$ and $D'$ do not depend on the undetermined parts of $W$.

To make $H'$ in \eqref{Hprimeform} we can add to $\tilde H$ a
kinetic and potential term involving outside oscillators only.  The
potential term (disregarding a constant) is
\be
f_{ij} \widetilde{x_i x_j} = \Tr f \widetilde{\xopsub}_\outout = 
\Tr f Q_\outfree \wopsub_\freefree Q_\outfree^T\,.
\ee
Since
\be
Q_\outfree = \sizedjoinmatrix(0, D)(\Nin,\Nin)(\Nout)\,,
\ee
this term is equivalent
to adding an arbitrary term to just the lower right part of $\tilde
K$.  Similarly the kinetic term is 
\be
g_{ij} \widetilde{P_i P_j} = \Tr g \widetilde{\popsub}_\outout = 
\Tr g Q^\invt_\freeout \pwopsub_\freefree Q^{-1}_\freeout\,.
\ee
Since
\be
Q^{-1}_\freeout =  \sizedstackmatrix(0, D')(\Nout)(\Nin,\Nin)\,,
\ee
this term corresponds to adding an arbitrary term to just the
lower right part of $\tilde T$.

That is to say we can write
\be[hprimeform]
H' = \half\Tr\left(T'\pwopsub_\freefree + K'\wopsub_\freefree\right)
\ee
with
\blea[tkform]
T' & = & {\arraycolsep 5pt \splitmatrix(I, 0, 0, ?)}\\
K' & = & \splitmatrix(K_\inin, K_\inout D, D^T K_\outin, ?)\,.
\elea

\subsection{Reduced constraints}

Now we rewrite our constraints, \eqsref{rhoconstraints}, in terms of the
$\bw$ coordinates.  We will keep only the parts of the constraint
equations that are not automatically satisfied because of the
ground-state modes.  For the expectation value constraints, 
from \eqsref{opistilde} and (\ref{eqn:expandoptilde}) we have
\blea
\<\xop\>_\outout & = & 
\widetilde{\xopexsub}_\outout = Q_\outfree\<\wop\>_\freefree Q_\outfree^T
+ \half Q_\outgs Q_\outgs^T\nonumber\\
&=& D \<\wop\>_\midmid D^T + \text{const}\\
\<\pop\>_\outout & = & \widetilde{\popexsub}_\outout
= {Q^{-1}_\freeout}^T \<\pwop\>_\freefree Q^{-1}_\freeout
+ \half Q^\invt_\gsout Q^{-1}_\gsout\nonumber\\
& = & D'^T \<\pwop\>_\midmid D' + \text{const}
\elea
where $\bw_\mid$ means the outside elements of $\bw_\free$, i.e.\
$w_{\Nin+1}\ldots w_{2\Nin}$.

These expectation value matrices must be the same in the vacuum as in
the vacuum-bounded state.  We can accomplish this by requiring that
$\<\pwop\>_\midmid$ and $\<\wop\>_\midmid$ are the same as in the
vacuum.  Our problem is now equivalent to one with only $\Nin$ outside
oscillators, and thus only $\Nin(\Nin+1)$ expectation value
constraints, regardless of the value of $\Nout$.

For the energy constraint, \eqref{Hrhoconstraint}, we are concerned
only with the renormalized energy $\Tr\rho H - \<0|H|0\>$.  Thus the
constant term in \eqref{tildeHdef} does not contribute, and we can
use $\tilde H$ from \eqref{tildeHdef-noconst}.  Once again there is no
dependence on $\Nout$.

\subsection{Derivation based on inside functions}

We would now like to redo the proceeding calculation in a way which
does not depend on the number of outside oscillators.  Then we can
remove the box from our system by taking $L \rightarrow\infty$ and $N
\rightarrow \infty$.  It appears that we have used the matrices $D$
and $D'$ which have an index that runs from $1$ to $\Nout$.  However,
we have used them only in particular combinations.  The quantities
which we need in our calculation are
\begin{enumerate}\parskip 0pt
\item $K_\inout \bar D$
\item $\bar D'\bar D$
\item $\bar D'\bar D'^T$
\item $\bar D^T K_\outout \bar D$.
\end{enumerate}
Each of these quantities is an $\Nin\times\Nin$ matrix, so it is
reasonable to imagine that they do not depend on $\Nout$ in the
$N\rightarrow\infty$ limit.

We proceed as follows:  From \eqsref{ddbardef} we have
\blea
\bar D &=& \xopexsub_\outin\\
\bar D' &=& \popexsub_\inout\,.
\elea
Any given element of $\xopex$ and $\popex$ has a smooth limit when
$\Nout$ is taken to infinity.  It is just a particular expectation
value of a half-line of coupled oscillators, which is a well-defined
problem.  We can express the above items in terms of such elements as
follows:
\begin{enumerate}
\item $K_\inout\bar D$ depends only on the first row of $\bar D$ which is
$\xopexsub_{\Ninn,\in}$ so it is already well-defined in the limit.

\item From \eqref{xoppinv} we have
\be[popxop]
\popex \xopex = \qtr I
\ee
so
$\bar D'\bar D = \popexsub_\inout \xopexsub_\outin = \qtr I -
\popexsub_\inin \xopexsub_\inin$, which does not depend on $\Nout$.

\item We expand $\popex \popex = \qtr Z^\invt Z^{-1} Z^\invt Z^{-1}$.  We
insert $I = Z \Omega_0 Z^T$ here to get
\be[kpop]
\popex \popex = \qtr Z^\invt \Omega_0 Z^{-1} = \qtr K\,.
\ee
Then $\bar D'\bar D'^T = \popexsub_\inout \popexsub_\outin
= \qtr K_\inin - \popexsub_\inin \popexsub_\inin$ which does not
depend on $\Nout$.

\item Using \eqsref{popxop} and (\ref{eqn:kpop}) we can write
\be
\xopex K = K \xopex = \popex \qquad\hbox{and }\qquad 
\xopex K \xopex = \qtr I
\ee
so
\bea
\qtr I &=& \xopexsub_\inout K_\outout \xopexsub_\outin
+ \xopexsub_\inin K_\inout \xopexsub_\outin\nonumber\\
&&+ \xopexsub_\inout K_\outin \xopexsub_\inin
+ \xopexsub_\inin K_\inin \xopexsub_\inin\nonumber\\
& = & \bar D^T K_\outout \bar D
+ \xopexsub_\inin (\popexsub_\inin - K_\inin \xopexsub_\inin)\nonumber\\
&&+ (\popexsub_\inin - \xopexsub_\inin K_\inin) \xopexsub_\inin
+ \xopexsub_\inin K_\inin \xopexsub_\inin
\eea
and so
\be
\bar D^T K_\outout \bar D = \qtr I
- \xopexsub_\inin \popexsub_\inin
- \popexsub_\inin \xopexsub_\inin
+ \xopexsub_\inin K_\inin \xopexsub_\inin
\ee
which does not depend on $\Nout$.

\end{enumerate}

Thus we can now take $N\rightarrow\infty$ with $\Nin$ fixed and have a
well-defined problem in terms of $\tilde K$ with a finite number of
free parameters.

\label{sec:endfirstreduction}

\subsection{Calculation of the reduced vacuum}

We are trying to compute $\tilde K$ in \eqref{ktildedef}
in the $\Nout\rightarrow\infty$ limit.  We will keep $\Nin$ and
the oscillator spacing $L_1 \equiv L/(N+1)$ fixed.
With the normalization in \eqref{Znorm} the vacuum normal mode matrix is
given by
\be
Z_{\mu\nu} = \sqrt{2\over N+1}{\sin k_\nu \mu\over\sqrt{\omega_\nu}}
\ee
with
\be
k_\nu = {\pi \nu\over N+1}
\ee
and
\be
\omega_\nu = {2 (N+1)\over L} \sin {k_\nu\over 2}
= {2 \over L_1} \sin {k_\nu\over 2}\,.
\ee
Thus
\bea
\xopexsub_{\mu\nu} & = & \half\left(Z Z^T\right)_{\mu\nu}
= {1\over N+1}\sum_{\alpha=1}^N {\sin k_\alpha \mu \sin k_\alpha \nu \over \omega_\alpha}\nonumber\\
& = & {L_1\over 2(N+1)} \sum_{\alpha=1}^N {\sin k_\alpha \mu \sin k_\alpha \nu\over \sin
(k_\alpha/ 2)}\,.
\eea
Now we use
\be[cosdif]
\cos(\theta-\phi) - \cos(\theta+\phi) = 2 \sin\theta \sin \phi
\ee
to write
\be[xopsum]
\xopexsub_{\mu\nu} = {L_1\over 4(N+1)}\sum_{\alpha=1}^N
   {\cos k_\alpha (\mu-\nu) - \cos k_\alpha(\mu+\nu) \over \sin(k_\alpha/2)}\,.
\ee
Using \eqref{cosdif} again, for any number $a$ we can write
\be
{\cos k_\alpha (a-1) - \cos k_\alpha a\over \sin(k_\alpha/2)} = 2 \sin k_\alpha (a-1/2)
\ee
and thus
\be[cossum]
{\cos k_\alpha (\mu-\nu) - \cos k_\alpha(\mu+\nu) \over \sin(k_\alpha/2)}
= 2 \sum_{a=\mu-\nu+1}^{\mu+\nu} \sin k_\alpha (a-1/2)\,.
\ee
If we put \eqref{cossum} into \eqref{xopsum} and bring the sum over $\alpha$
inside the sum over $a$ we get a sum that we can do,
\be
2 \sum_{\alpha=1}^N \sin {\pi \alpha (a-1/2) \over N+1}
= \cot {\pi(2a-1)\over 4(N+1)} + (-1)^a\,.
\ee
Now we sum this over $a$.  Since $\mu-\nu$ and $\mu+\nu$ have the same parity,
the $(-1)^a$ term does not contribute and we get
\be
\xopexsub_{\mu\nu} = {L_1\over 4(N+1)} \sum_{a=\mu-\nu+1}^{\mu+\nu} 
    \cot {\pi(2a-1)\over 4(N+1)}\,.
\ee

The sum over $N$ has been eliminated.  In the $N\rightarrow\infty$
limit, the argument of $\cot$ goes to zero and so we can
use $\cot x = 1/x + O(x^{-3})$ to get
\be
\xopexsub_{\mu\nu} = {L_1 \over \pi} \sum_{a=\mu-\nu+1}^{\mu+\nu} {1\over 2a-1}\,.
\ee
The sum can be done using special functions:
\be[finalxop]
\xopexsub_{\mu\nu} = {L_1 \over 2\pi}
\left[\psi\left(\mu+\nu+\half\right) - \psi\left(\mu-\nu+\half\right)\right]
\ee
where $\psi$ is the digamma function, $\psi(x) = \Gamma'(x)/\Gamma(x)$.

\label{sec:xopexform}

To compute $\popex$ we write the inverse of the normal mode matrix,
\be
Z^{-1}_{\mu\nu} = \sqrt{2\omega_\mu\over N+1}\sin k_\mu \nu
\ee
and
\bea
\popexsub_{\mu\nu} & = & \half\left(Z^\invt Z^{-1}\right)_{\mu\nu}
 = {1\over N+1}\sum_{\alpha=1}^N\omega_\alpha\sin k_\alpha \mu \sin k_\alpha \nu\nonumber\\
& = & {2\over L_1(N+1)} \sum_{\alpha=1}^N\sin {k_\alpha\over 2}\sin k_\alpha \mu \sin k_\alpha \nu\,.
\eea
In this case the sum can be done directly.  Using \eqref{cosdif} we can write
\be
\popexsub_{\mu\nu} = \popone_{\mu-\nu} - \popone_{\mu+\nu}
\ee
where
\bea
\popone_\lambda & \equiv & {1\over L_1(N+1)} \sum_{\alpha=1}^N \sin {k_\alpha\over 2} \cos k_\alpha \lambda\nonumber\\
& = & {1\over 4 L_1 (N+1)}
\left[\cot{\pi(2\lambda+1)\over 4(N+1)}
 - \cot{\pi(2\lambda-1)\over 4(N+1)} - 2 (-1)^\lambda\right]\,.
\eea
Again we take $N \gg \lambda$ to get
\be
\popone_\lambda = {1\over \pi L_1}\left({1\over 2\lambda+1}
 - {1\over 2\lambda-1}\right)\\
= - {2\over \pi L_1(4\lambda^2-1)}
\ee
and
\be[finalpop]
\popexsub_{\mu\nu} = {2\over \pi L_1}
\left({1\over 4(\mu+\nu)^2-1} - {1\over 4(\mu-\nu)^2-1}\right)\,.
\ee
\Eqsref{finalxop} and (\ref{eqn:finalpop}) give $\xopex$ and $\popex$
in the $\Nout\rightarrow\infty$ limit.  Using these values in the
procedure of sections \ref{sec:startreduction} through
\ref{sec:endfirstreduction} we can compute the matrix $\tilde K$
numerically for a system with inside length $\Lin$ but no outside box.

\label{sec:endreduction} 

\section{Numerical computation of the reduced vacuum}
\label{sec:numer}

We have computed numerically the reduced coupling matrix $\tilde K$,
by following the procedure of
\secsref{startreduction}--\ref{sec:endreduction}.
This is a fairly straightforward problem in numerical analysis; the
number of steps grows as $\Nin^3$.  However, in order to produce
accurate results for $\Nin \agt 6$ it is necessary to use very
high-precision floating-point numbers.  The necessary number of bits
of mantissa in the representation appears to be about $10\Nin$.  The
code was written in Lisp and executed on DEC$^{\scriptscriptstyle\text{TM}}$
Alpha$^{\scriptscriptstyle\text{TM}}$ workstations.

The resulting matrix $\tilde K$ is a tridiagonal matrix which gives a set of
self-couplings and nearest-neighbor couplings for the fictitious
oscillators $\bw_\free$.  We can express these couplings as multiples
of the couplings for a regular chain of $\Nfree$ oscillators with
spacing $L_1$.  Thus we write the self-coupling as
\be
\tilde K_{\mu\mu} = -g f_\mu\qquad \left(\notsum_\mu\right)
\ee
and the nearest-neighbor coupling as 
\be
\tilde K_{\mu,\mu+1} = 2g f_{\mu+1/2} \qquad \left(\notsum_\mu\right)\,.
\ee

These coupling coefficients converge rapidly to a universal form
$f(x)$ where $L_1 \mu\rightarrow x$ in the continuum limit.  Some
results are shown in \mypsfigr{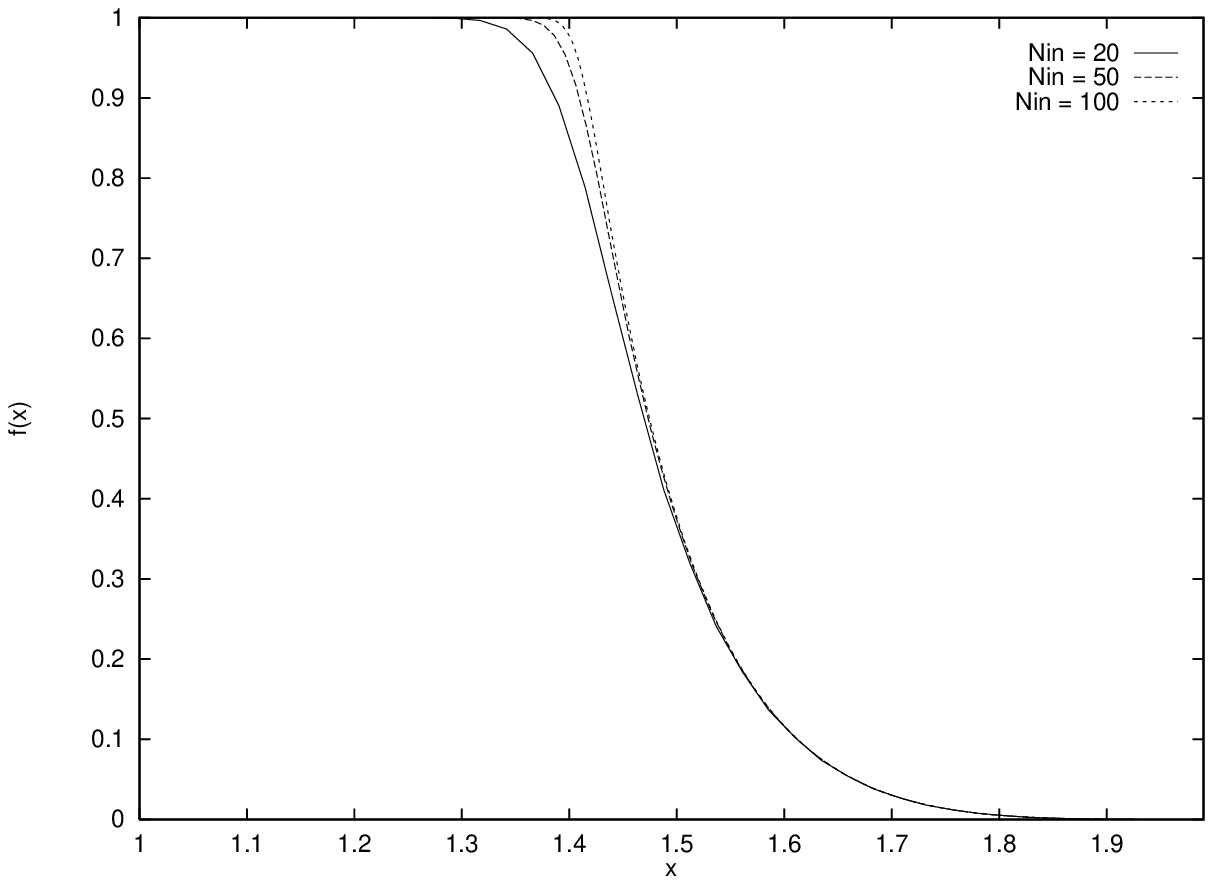}{The ratio of the coupling
coefficients in the reduced problem to what they would be in a regular
problem of $\Nfree$ oscillators, plotted against $x = L_1\mu$, for
$\Lin = 1$}.  We can see that $f(x) \approx 1$ until $x\sim 1.4$ at
which point it begins to fall and asymptotically approaches $0$ as
$x\rightarrow 2$.  For values of $x$ near $2$, $f(x)$ is well fit by
\nopagebreak
\be[fxform]
f(x) = a (2-x)^4
\ee
with $a\approx 3.2$, as shown in \mypsfigr{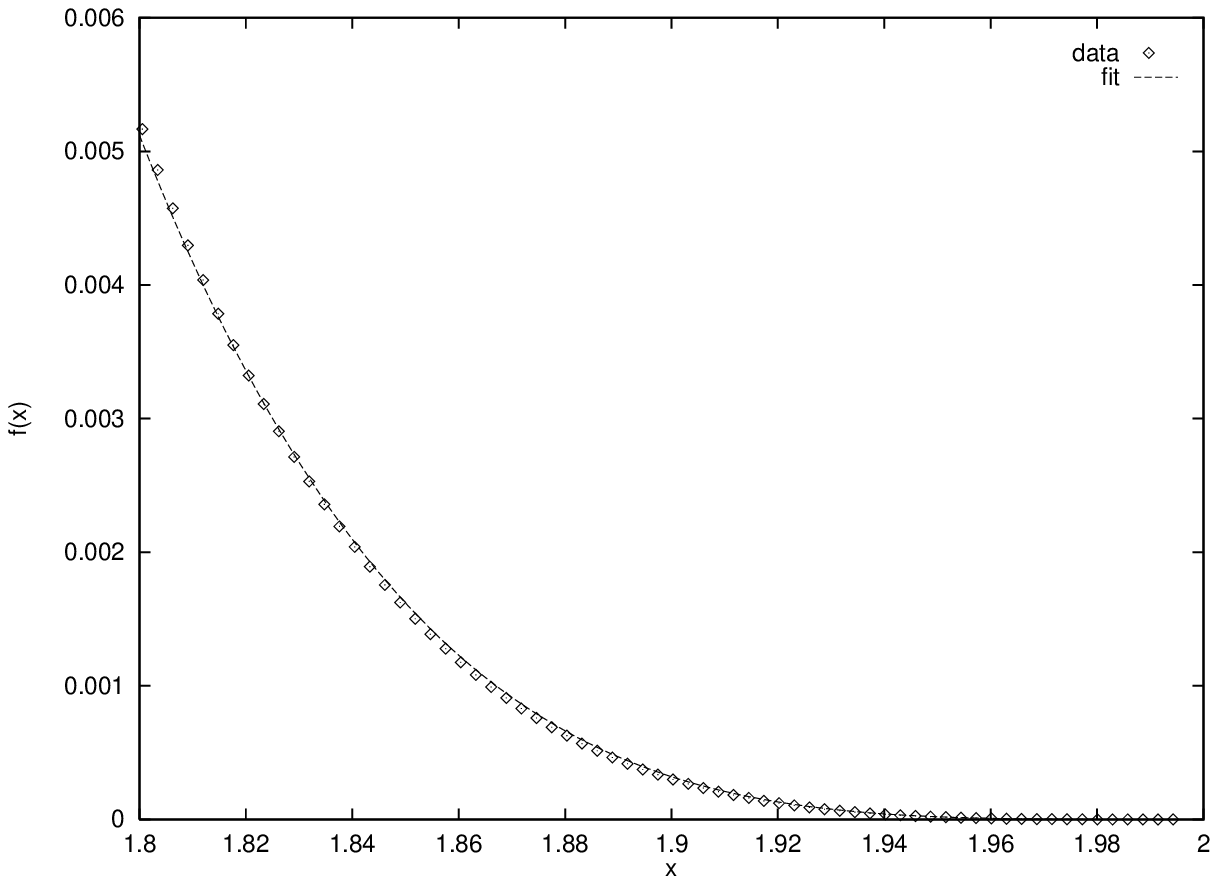}{The
coupling coefficient ratio for $\Nin=175$ in the region $x>1.8$ and
the fit $f(x) = 3.2 (2-x)^4$}.

\label{sec:smallfreqs}
Some typical normal modes of the reduced vacuum are shown in
\mypsfigroff{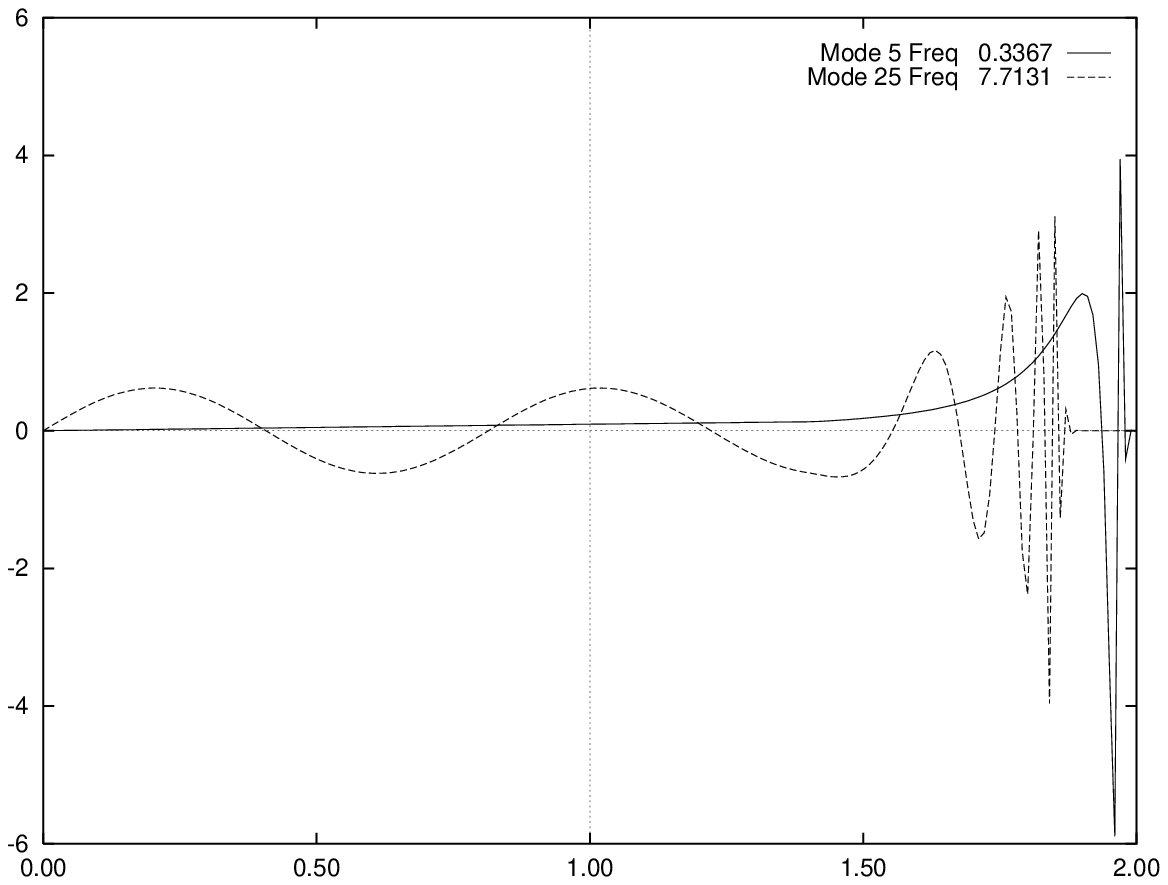}{The $5^{\protect\text{th}}$ and
$25^{\protect\text{th}}$ normal modes in the reduced vacuum, computed
with $\Nin=175$}.  They
are sine waves while $f(x) \sim 1$ and then begin to oscillate faster
and faster as $f(x)$ shrinks.  At first the amplitude of the
oscillations grows but for larger $x$ it shrinks rapidly to zero.  The
wavenumbers in the inside region (and thus the frequencies) are
smaller than we would find for a rigid box because most of the
oscillations are in the part of the outside region where $f(x) \ll 1$.
In fact, as $\Nin\rightarrow\infty$ we would expect the low-lying
frequencies to go to zero, for the following reason.

We can find the frequencies by computing the normal modes of a Hamiltonian
\be
H = \int_0^L dx dy \left(T(x-y)\pi(x)\pi(y)+K(x-y)\phi(x)\phi(y)\right)\,,
\ee
which requires solving the eigenvector equation
\be[generaldiffeq]
\int_0^L dy dz T(x-y) K(y-z) g(z) = \lambda g(x)
\ee
with the boundary conditions 
\blea[generalbcs]
g(0) &=& 0\\
g(L) &=& 0\label{eqn:rightbc}\,.
\elea
Since \eqref{generaldiffeq} is a second-order differential equation we
expect two degrees of freedom in the solution.  However, one degree of
freedom is manifestly the overall scale, which does not affect the
boundary conditions.  Since there are two boundary conditions but only
one free parameter, we can expect to find solutions only for
particular values of $\lambda$.  For example, for the usual scalar
field Hamiltonian the general solution to \eqref{generaldiffeq} would be
\be
g(x) = c \sin(\sqrt{\lambda} x + \delta)\,.
\ee
To satisfy \eqsref{generalbcs} we need to choose $\delta=0$ and
$\sqrt{\lambda} = n\pi/L$ for some integer $n$.

However, if we use
\be[redvacH]
H = \half \int_0^{2\Lin} f(x)\left[\pi(x)^2 
 + \left({d\phi\over dx}\right)^2\right] dx\,,
\ee
with $f(x)\rightarrow a(2-x)^4$, as suggested by
\figref{vacuum-freq-fit}, we will get a continuum of frequencies.  The
problem is that since $f(x)\rightarrow 0$ as $x\rightarrow 2$ the
boundary condition there does not really constrain $g(x)$.  There can
be arbitrary changes in $g(x)$ near $x=2$ and so \eqref{rightbc}
can always be satisfied.  Since there is only one effective boundary
condition and one effective degree of freedom, we expect to be able
to find a solution for any $\lambda$.  Thus in the continuum limit
there are modes with arbitrarily low frequencies.  This is not an
unreasonable conclusion, since although the range of $x$ is finite, we
are using it to represent the infinite half-line.  In the infinite
vacuum there is no right-hand boundary condition, and
there are modes of every frequency.

This conclusion is confirmed by numerical results. In 
\mypsfigr{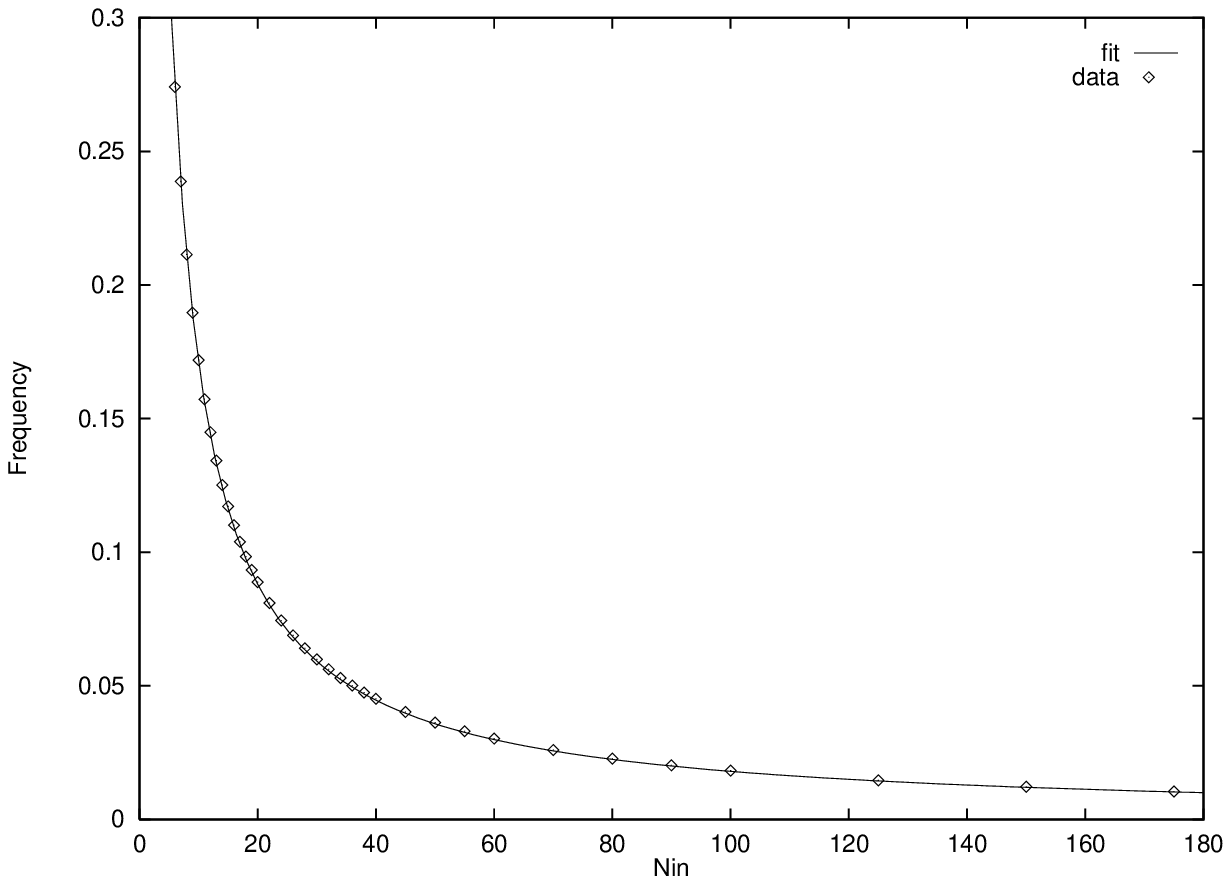}{The lowest frequency of the reduced
vacuum and the fit $1.81 x^{-1} - x^{-2}$} we plot
the lowest normal-mode frequency versus $\Nin$.  As shown in the
figure, the frequencies are well fit by a curve
\be
a x^{-1} - x^{-2}
\ee
with $a\approx 1.81$.  If this form is correct, in the $\Nin
\rightarrow\infty$ limit the lowest frequency goes to zero.

If we go to a vacuum-bounded state we will introduce some finite
temperature.  We then expect that the non-zero temperature will
increase the frequencies in such a way that there are only a finite
number of low-lying modes and thus a finite entropy.  However, in the
limit where $T\rightarrow 0$ we do expect an entropy-to-energy
relation equivalent to a system with infinitesimal frequencies.  We
now consider this situation.

\section{The low-energy limit}
In the low-energy regime we can make a first-order expansion around
the vacuum.  To do this we note that the only dependence on $\beta$ in
our equations (see \oldpaper) is through expectation values that
depend on $\coth(\beta\omega_\alpha/2)$.  For large $\beta$ we can
approximate
\be
\coth{\beta\omega_\alpha\over 2} \approx 1+2e^{-\beta\omega_\alpha}\,.
\ee
The change in $\coth(\beta\omega_\alpha/2)$ is the largest for the
smallest frequency, which we will call $\omega_1$.  We will ignore
$e^{-\beta\omega_\alpha}$ for larger $\omega_\alpha$ by comparison
with $e^{-\beta\omega_1}$.  Thus we take
\blea[loweapproxfirst]
\ddelta\coth{\beta\omega_1\over2} &=& 
2e^{-\beta\omega^\vac_1} \equiv 2\epsilon \\
\ddelta\coth{\beta\omega_\alpha\over 2} &=& 0 \qquad\hbox{for $\alpha> 1$}\,.
\elea
Then we write
\blea
T'_\midmid &=& \tilde T_\midmid + \ddelta T_\midmid\\
K'_\midmid &=& \tilde K_\midmid + \ddelta K_\midmid
\elea
where $\ddelta K_\midmid$ and $\ddelta T_\midmid$ are $O(\epsilon)$.
These changes give rise to $O(\epsilon)$ changes in $U$ and
the $\omega_\alpha$, which in turn give rise to $O(\epsilon)$ changes in
$\<w_m w_n\>$ and $\<P_{w_m} P_{w_n}\>$.
\abovedisplayskip 10pt plus3pt minus5pt
\belowdisplayskip \abovedisplayskip

Since overall $\<w_m w_n\>$ and $\<P_{w_m} P_{w_n}\>$ cannot change we
must have
\blea[loweapproxlast]
0 = \ddelta\<w_m w_n\> = \sum_\alpha\bigg(&&
-{\ddelta\omega_\alpha\over 2\omega_\alpha^2} U^\vac_{m\alpha} U^\vac_{n\alpha}
+ {1\over 2\omega^\vac_\alpha} \ddelta U_{m\alpha} U^\vac_{n\alpha}
+ {1\over 2\omega^\vac_\alpha} U_{m\alpha} \ddelta U_{n\alpha}\bigg) \nonumber\\
&&+ {1\over \omega^\vac_1} U^\vac_{m1} U^\vac_{n1} \epsilon\\
0 = \ddelta\<P_{w_m} P_{w_n}\> = \sum_\alpha\bigg(&&
{\ddelta\omega_\alpha\over 2}\uvacinv_{\alpha m} \uvacinv_{\alpha n} 
+ {\omega^\vac_\alpha\over 2}\ddelta U^{-1}_{\alpha m} \uvacinv_{\alpha n}
+{\omega^\vac_\alpha\over 2}\uvacinv_{\alpha m}\ddelta U^{-1}_{\alpha n}\bigg)
\nonumber\\
&&+ \omega^\vac_1 \uvacinv_{m1} \uvacinv_{n1}\epsilon\,.
\elea

We thus have $\Nin(\Nin+1)$ linear equations for $\Nin(\Nin+1)$
unknown values of $\ddelta T_\midmid$ and $\ddelta K_\midmid$, which are
readily solved.  Since the inhomogeneous part of these equations is
$O(\epsilon)$, all the results must be $O(\epsilon)$ as well.  In
particular, the $\ddelta\omega_\alpha$ are $O(\epsilon)$.  Now
if $T$ is very small as compared to all the $\omega_\alpha$, then
$\epsilon$ will be small as compared to all the parameters of the
problem, and so the first-order approximation will be good.  For any
fixed number of oscillators $\Nin$ there will be some minimum
frequency $\omega_1$, and if we take $\beta\ll 1/\omega_1$ we will
always be in this regime.

Now the entropy $S$ depends only on $\beta$ and the $\omega_\alpha$.
Since the modes are uncoupled,
\be
S = \sum_\alpha S_1(\beta\omega_\alpha)
\ee
with
\be
S_1(\beta\omega) = -\ln(1-e^{-\beta\omega})
 + {\beta\omega\over e^{-\beta\omega} -1}\,.
\ee
Since $\beta\omega_\alpha \gg 1$ all the terms are very small, and the
$\omega_1$ term dominates,
\be
S \approx S_1(\omega_1) = (1+\beta\omega_1)e^{-\beta\omega_1} +
O(e^{-2\beta\omega_1})\,.
\ee
Since $\epsilon$ drops exponentially with increasing $\beta$, we
expect that for $\beta$ large enough, $\beta\ddelta\omega_1 \ll 1$ so
that
\be[loweS]
S = (1+\beta\omega^\vac_1)e^{-\beta\omega^\vac_1}+O(\epsilon^2)\,.
\ee
The value of $S$ given in \eqref{loweS} is the one we would get
from a rigid box with length
\be[lowelinprimedef]
\Linp = \pi/\omega^\vac_1\,.
\ee

To approximate the energy, we proceed along the lines of
\oldsecref{IX}.  The direct calculation is made difficult by the fact
that, while $H'$ differs from $H$ only by $O(\epsilon)$, we must
subtract from both Hamiltonians a large ground-state energy.  Instead
we work by integrating on $T$.  From $F=E-TS$ and $dF= -S dT$ we find
\be[loweint]
E(T) = TS(T) - \int_0^T S(T') dT'\,.
\ee
Now $\Linp$ depends only on $\omega_1^\vac$, which depends on $\Nin$
but not on $\beta$.  If \eqref{loweS} is valid for a particular $\Nin$
at $\beta=1/T$ is it valid for $T'<T$ and $\beta'=1/T'>\beta$.  
Thus both $S(T)$ and $S(T')$ in \eqref{loweint} are just the entropy of a
rigid box of length $\Linp$.  Thus the entropy-to-energy
relationship is just $S(E) = S^\rb(\Linp;E)$, the entropy as a
function of the given energy in a rigid box with length
$\Linp$.

For such a rigid box at very low energy we find
\be
E={\omega^\rb_1\over e^{\beta\omega^\rb_1} -1}\approx \omega^\rb_1
e^{-\beta\omega^\rb_1}
\ee
and thus
\be
S = \left(1+\ln{\omega^\rb_1\over E}\right){E\over\omega^\rb_1}
\ee
where $\omega^\rb_1 = \pi/\Linp = \omega^\vac_1$ is the frequency of
the lowest mode.

Now for any given $\Nin$ we get some $\omega^\vac_1$.  As discussed in
\secref{smallfreqs}, the larger $\Nin$ we choose, the smaller
$\omega^\vac_1$ we will have.  For $\Nin$ fixed we can choose
$\beta\gg1/\omega^\vac$ and proceed as above to get a large value of
$\Linp$.  However, we are really interested in the continuum limit at
fixed temperature.  If we increase $\Nin$ with $\beta$ fixed we will
find that $\omega_1$ (and eventually an arbitrary number of the
$\omega_\alpha$) will become smaller than $1/\beta$.  When this
happens, the approximations of Eqs.\
(\ref{eqn:loweapproxfirst}--\ref{eqn:loweapproxlast}) will no longer
be good.

However, we do not expect the entropy to decrease drastically in this
limit.  To make the entropy small would require making all the
frequencies large.  If the frequencies were large, the approximations
we have used would again become valid.  Then we could argue as before
that the entropy should be large.  It would be hard to have a
consistent picture.

Now consider the limit as $T\rightarrow 0$.  For each $T$ we start
with some initial number of oscillators $\Nin^{(0)}$.  We choose
$\Nin^{(0)}$ not too large, such that $\omega^\vac_1 \gg T$.  With
this value of $\Nin$, we find $\Linpsup{0} \sim \pi/\omega^\vac_1$.
We then let $\Nin\rightarrow\infty$ and we conjecture that the entropy
does not change much, and thus in the continuum limit $S(E) \sim
S^\rb(\Linpsup{0}; E)$.  As we decrease $T$ we can decrease the
initial $\omega^\vac_1$ and so increase $\Linpsup{0}$ without bound.
Thus we make the following conjecture:
\begin{quotation}
For a given energy $E$, let $\Linp(E)$ be the length of a rigid
box such that the vacuum-bounded state with energy $E$ and length
$\Linp$ has entropy $S(E) = S^\rb(\Linp(E);E)$.  Then
\be
\lim_{E\rightarrow 0} {\Linp(E)\over \Lin} = \infty\,.
\ee
\end{quotation}

To support this conjecture numerically we turn to direct calculation
of energy and entropy values for vacuum-bounded states at low
temperature.  For various fixed values of $\beta=1/T$ and for various
numbers of oscillators we compute $S$ and $E$ and from them the
equivalent length $\Linp$.  Some results are plotted in
\mypsfigr{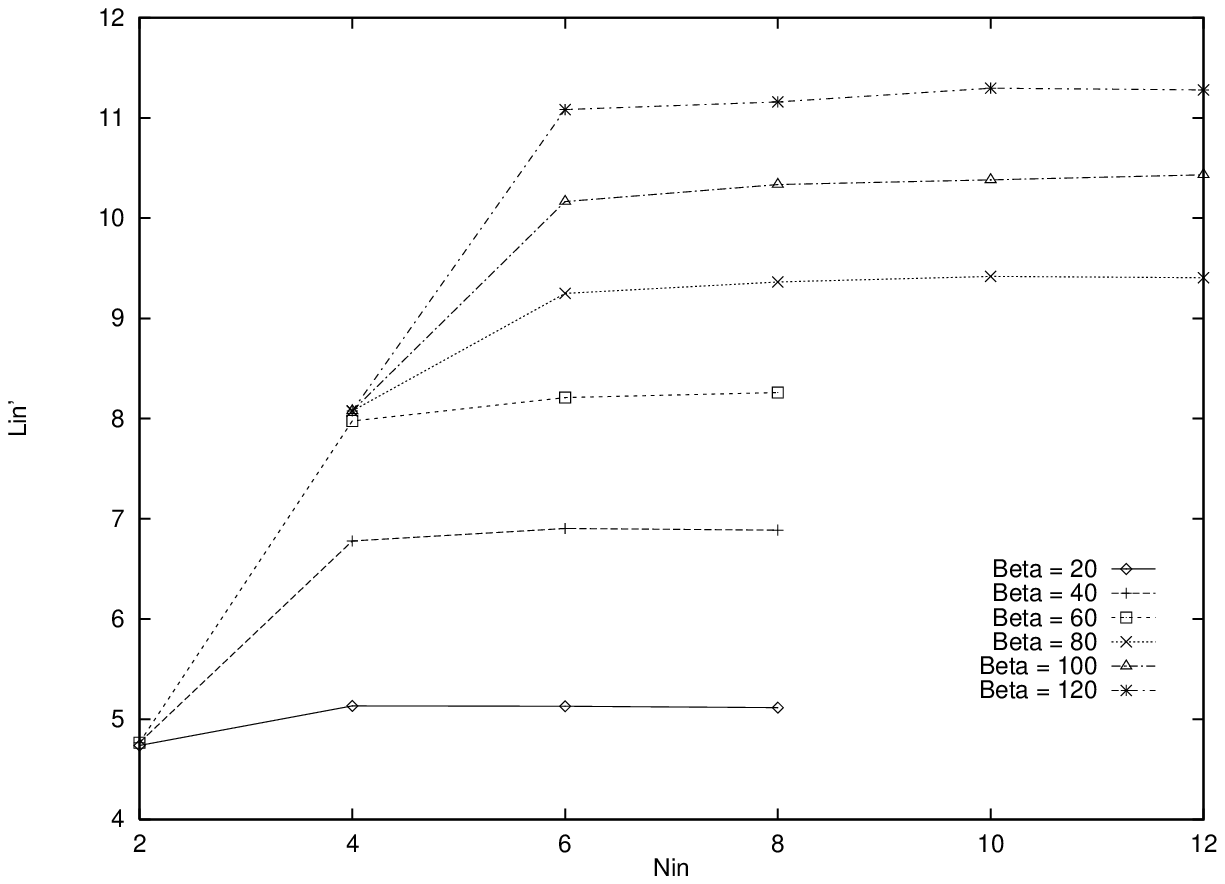}{The length $\Linp$ of a rigid box that gives
the same $S(E)$ as a vacuum-bounded state at temperature $T=1/\beta$
and $\Lin = 1.0$}.  While $\Nin$ is still small enough for the
approximations of Eqs.\
(\ref{eqn:loweapproxfirst}--\ref{eqn:loweapproxlast}) to be valid,
$\Linp$ grows with $\Nin$.  Once $\Nin$ has left this regime, it
appears that $\Linp$ levels off.  It is at least reasonable to believe
that there is no further change in $\Linp$ as $\Nin\rightarrow\infty$.
In \mypsfigr{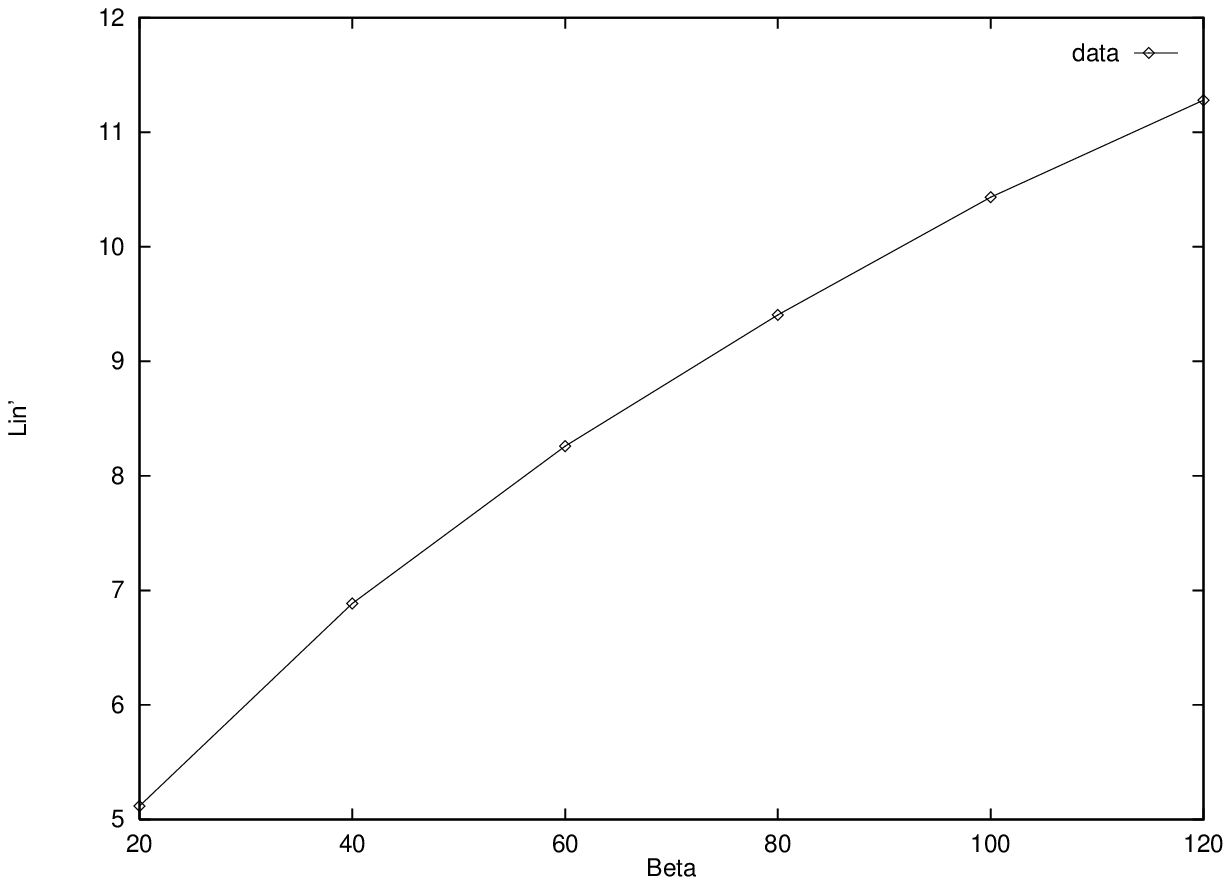}{The length $\Linp$ of a rigid box
with the same $S(E)$ plotted against $\beta$.  Each point is the value
for the largest number of oscillators available} we plot the eventual
level of $\Linp$ versus $\beta$.  It appears that the limiting value
of $\Linp$ grows nearly linearly with $\beta$, and thus
$\Linp\rightarrow\infty$ as $E\rightarrow 0$ as conjectured.

\section{High Energies}
In this section we use the results of \secref{reduction} to redo the
computation from \oldpaper\ of the bound on the entropy of a
high-energy vacuum-bounded state.  The difference here is that $\Lout$ and
thus $\Nout$ are explicitly infinite, whereas in \oldpaper\ we were
limited to $\Nout \sim \Nin$.

First we solve numerically for the maximum-entropy vacuum-bounded
state.  We first compute the reduced vacuum as in \secref{numer} and
then we find the vacuum-bounded state, exactly as in \oldsecref{VII},
with the vacuum couplings given by $\tilde K$.  The results are shown
in \mypsfigr{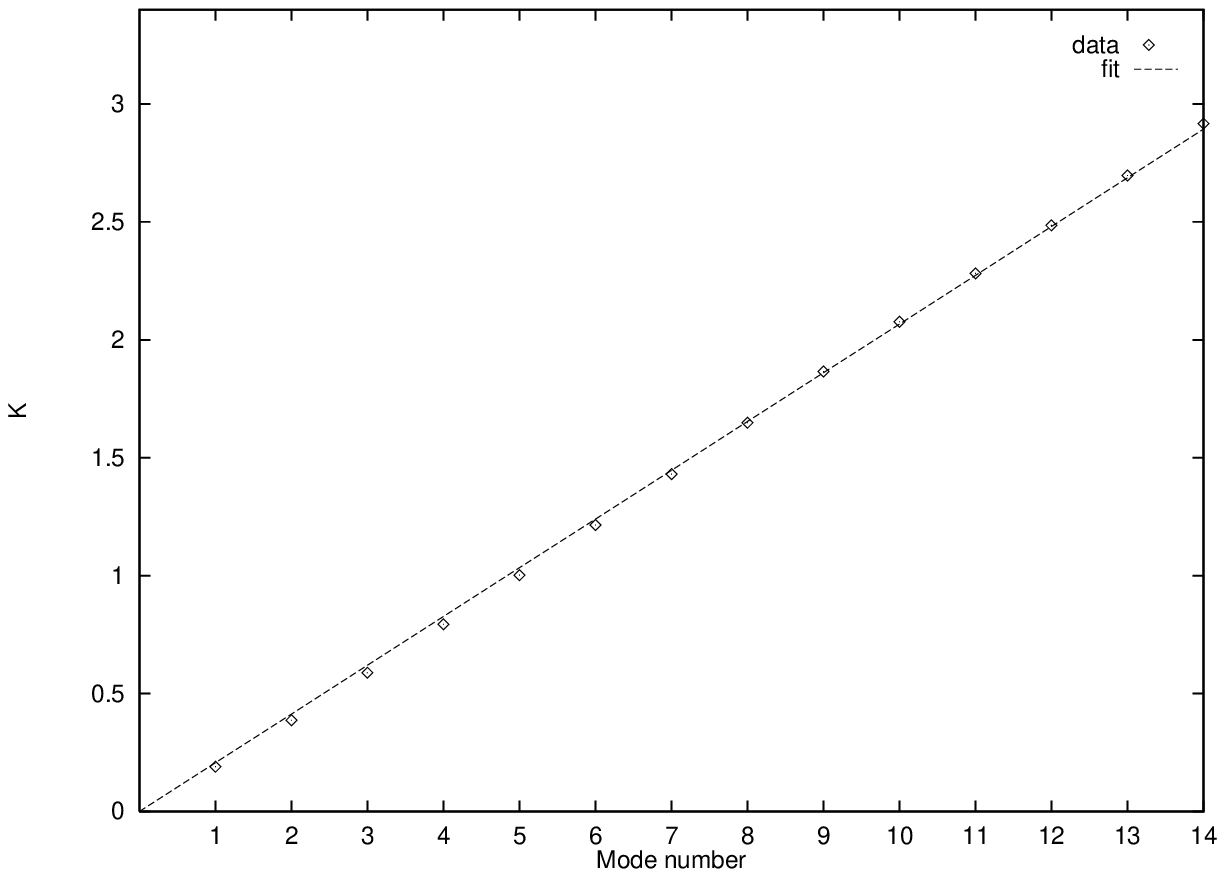}{The numerically computed wavenumbers
compared with the best-fit line through the origin for $\Lin=1.0$,
$\beta=0.5$, $\Nin=12$}.  As in \oldpaper\ we find that the mode number
spacing is very nearly uniform.

Using this as an ansatz, we can repeat the calculation of
\oldsecref{VIII}.  We will work in the $\bw$ coordinates, and set
\nopagebreak
\be
\<w_\Ninn^2\> = \<w_\Ninn^2\>^\vac\
\ee

\subsection{The vacuum}\label{sec:vacuum-calc}

In \secref{xopexform} we computed the values of $\xopex$.  To
convert to $\bw$ coordinates we proceed as follows:  From 
\secref{xfromw}, $\bx = Q \bw$ and so
\be[xfromw]
\bx_\out = R \bw_\out = D \bw_\mid + Y_\outgs \bw_\gs\,.
\ee
In the numerical work we found that with our choice for $D$ we got
\be
\tilde K_\inmid = \lowerleftnonzero{-g}\,.
\ee
Since
\be
\tilde K_\inmid = K_\inout D = \lowerleftnonzero{-g} D
\ee
it follows that the first row of $D$ is $(1,0\ldots 0)$.  Thus from 
\eqref{xfromw} we get
\be
x_\Ninn = w_\Ninn + (Y_\outgs)_1 \cdot \bw_\gs
\ee
where $(Y_\outgs)_1$ denotes the first row of $Y_\outgs$.
Now $\widetilde\wopsub_\gsgs = (1/2) I$ and $\widetilde\wopsub_\gsfree = 0$,
so
\be
\<x_\Ninn^2\> =  \<w_\Ninn^2\> + \half\left(Y_\gsout Y_\gsout^T\right)_{11}\,.
\ee
Since the last term is non-negative, we have
\be
\<w_\Ninn^2\>^\vac \le \<x_\Ninn^2\>^\vac\,.
\ee
This gives us an only upper bound on $\<w_\Ninn^2\>^\vac$, but that is
sufficient to produce an upper bound on $S(E)$, which is what we are
looking for.

From \eqref{finalxop} we have
\be
\<x_\Ninn^2\>^\vac = \xopexsub_{\Ninn,\Ninn}
 = {L_1 \over 2\pi} \left[\psi\left(2\Nin + {5\over 2}\right)
- \psi\left(\half\right)\right]\,.
\ee
We would like to evaluate this expression in the
$\Nin\rightarrow\infty$ limit with $\Lin$ fixed.  There is a prefactor
of $L_1$, which goes to zero in this limit, but that is just an
artifact of the conventions we have used for the discrete problem, and
will appear in the finite-energy vacuum-bounded states as well.  For
large $x$,
\be[digamma-big]
\psi(x) \sim \ln x + O(1/x)\,,
\ee
so without the prefactor there is a logarithmic divergence.  We are
interested in the $\ln \Nin$ term, and in the constant term, but we
will ignore any terms of order $1/\Nin$ or lower.

We use \eqref{digamma-big} and $\psi(1/2) = -\gamma - 2\ln 2$, where
$\gamma$ is Euler's constant, to get
\be
\<x_\Ninn^2\>^\vac = {L_1\over 2\pi}\left[\ln 2\Nin + \gamma + 2\ln 2
+ O\left({1\over \Nin}\right)\right]
\ee
and so
\be[wwvacfinal]
\<w_\Ninn^2\>^\vac \le \<x_\Ninn^2\>^\vac
 = {L_1 \over 2\pi}\left[\ln 8\Nin+\gamma+O{\left(1\over \Nin\right)}\right]\,.
\ee

\subsection{Evenly spaced wavenumbers}
The calculation of \oldsecref{VIII B} does not have any depedence on
$N$, except in the factor $L/N$ which here is $L_1$.  Thus we can use
the same procedure with $\bw$ in place of $\bx$, but the same mode
function, to find (see \oldpaper, Eqs. (8.33) and (8.37))
\be
\<w_\Ninn^2\> \gtrsim 
 {L_1\over2 \pi}\left(\ln{8\Nnorm\sin\pi\Delta\over\pi}+\gamma
+2\pi\tau'\Delta+\ln{1-e^{-4 \pi\tau' \Delta}\over 4 \pi\tau' \Delta}
\right)\,.
\ee

Setting $\<w_\Ninn^2\>$ = $\<w_\Ninn^2\>^\vac$ from \eqref{wwvacfinal}
and using the same approximations and in \oldpaper\ we find
\be
\Linp \le \Lin + {1\over 2\pi T} \ln {\Lin T} +O\left({1\over T}\right)
\ee
just as in \oldpaper.  This confirms our conclusion that the
difference in entropy is bounded by
\be
\ddelta S(E) \lesssim {1\over 6}\ln\Lin T \approx {1\over 6}\ln S^\rb\,,
\ee
and the conclusions of \oldpaper\ remain unchanged.

\section{Discussion}
We have improved the calculations of \oldpaper\ by handling an infinite
number of outside oscillators with only a finite number of degrees of
freedom.  This is possible because the total number of modes that can
be excited depends only on the number of inside oscillators.  Our
technique is in some ways similar to representing an infinite
half-line by a finite number of oscillators with larger and larges
spaces between them.

Using this technique we look at very low energy states which look like
the vacuum except in a particular region.  We find that, because of
very low frequencies in the vacuum state, much more entropy can be
stored in a small region with the vacuum-bounded condition than could
be stored in such a region with a rigid boundary.  To quantify the
difference we let $\Linp$ be the size of a rigid-bounded system
with the same entropy and energy as a vacuum-bounded system of length
$\Lin$.  We argue that one should expect the ratio $\Lin'/\Lin$ to
grow without bound as the energy decreases.  Numerical calculations
lend support to this argument.

We also redo the high-energy calculation from \oldpaper\ to reach the
same conclusion with fewer approximations.

\section{Acknowledgments}

The author would like to thank Alan Guth for much advice and
assistance; Leonid Levitov, Edward Farhi and Andy Latto for helpful
conversations; Harlequin Inc.\ for providing a copy of their LispWorks
product on which some of the computations were done; Bruno Haible and
Marcus Daniels for their CLISP Common Lisp implementation; and Kevin
Broughan and William Press for making available a version of {\em
Numerical Recipes} translated into LISP\footnote{These routines and
many others are now available on a
CDROM\protect\cite{numrecip:cdrom}}.

Portions of this work are reprinted from \cite{olum:thesis}, copyright
Massachusetts Institute of Technology, with permission.

This work was supported in part by funds provided by the
U.S. Department of Energy (D.O.E.) under cooperative research
agreement DE-FC02-94ER40818 and supported in part by the National
Science Foundation.



\end{document}